\documentclass[a4paper,11pt]{article}
\usepackage{pos}
\usepackage{cleveref}

\title{Parton and nuclear dynamics at large $x$: the CJ26 global QCD analysis}
\author*[a,b]{Alberto Accardi}
\author[c]{Matteo Cerutti}
\author[b,d]{Cynthia E. Keppel}
\author[e]{Shujie~Li}
\author[f]{J.~F.~Owens}
\author[b]{Sanghwa Park}
\author[b]{Peter Risse}

\affiliation[a]{Christopher Newport University,
Newport News, Virginia 23606, USA}

\affiliation[b]{Jefferson Lab, 
Newport News, Virginia 23606, USA}

\affiliation[c]{Universit\'e Paris-Saclay – CEA – IRFU,
91191 Gif-sur-Yvette, France}

\affiliation[d]{3q~Scientific~Consulting,~LLC, 
Newport News, Virginia 23602, USA}

\affiliation[e]{Lawrence Berkeley National Laboratory, Berkeley, California 94720, USA}
\affiliation[f]{Florida State University, Tallahassee, Florida 32306, USA}

\emailAdd{accardi@jlab.org}

\abstract{We present the new CTEQ-JLab global QCD analysis, called CJ26, that simultaneously fits parton distribution functions of the proton and nuclear effects in deuteron targets. This work includes, for the first time in a global fit, the full set of JLab 6 GeV inclusive DIS measurements and the first published JLab 12 GeV data. The analysis focuses on the large-$x$ region, and carefully treats the interplay between higher-twist effects and off-shell corrections in low-energy proton and deuteron target data. We discuss their effect on quark and gluon distributions, and highlight the importance of quantifying the experimental correlated systematic uncertainty to achieve fits of the highest accuracy. We also  outline how upcoming data may help unravel the nuclear and partonic dynamics in light nuclei. The extracted next-to-leading order parton distributions are available on the CJ web pages in LHAPDF format for general use, accompanied by leading- and higher-twist structure functions calculated with the simultaneously fitted PDFs, higher-twist corrections, and off-shell corrections. This write up clarifies several points that became clearer while finalizing the CJ26 paper and can be read as a companion to the latter.}

\renewcommand{\hookAfterAbstract}{%
\vskip1cm\par\bigskip
\textsc{Preprint number}: JLAB-THY-26-4991}

\FullConference{The 33rd International Workshop on Deep Inelastic Scattering and Related Subjects (DIS2026)\\
4 - 8 May 2026\\
Bologna, Italy\\}

\begin{document}
\maketitle

\section{Introduction}

Parton distribution functions (PDFs) extracted from Deep Inelastic Scattering (DIS) and other high-energy proton collisions are usually thought of as leading-twist quantities whose scale dependence is described by perturbative QCD. At large Bjorken $x$, however, this picture becomes increasingly incomplete. The available DIS data move toward lower $Q^2$, target-mass effects become important, and power-suppressed contributions can be numerically large. Moreover, the flavor separation of up and down quark PDFs requires deuteron target data, introducing nuclear corrections that are themselves correlated with the partonic quantities being extracted. A global analysis that simply removes low-$Q^2$ data gives up a substantial amount of information on the valence region.

The CTEQ-JLab (CJ) program was developed to make use of this information while treating the relevant corrections within the global fit. Earlier analyses established a framework in which target-mass corrections, residual power corrections, and nuclear effects in the deuteron are fitted together with the PDFs \cite{Accardi:2016qay}. The CJ22 analysis subsequently incorporated new electroweak constraints on the light-quark distributions \cite{Accardi:2023gyr}, while the CJ22ht study investigated in detail how the implementation of higher-twist (HT) corrections affects the extraction of large-$x$ PDFs and nuclear quantities \cite{Cerutti:2025yji}.

The new CJ26 analysis \cite{Accardi:2026hdv} takes the next step by including, for the first time in a global QCD fit, the complete set of available Jefferson Lab 6~GeV inclusive DIS measurements together with the first published JLab 12~GeV data. The resulting global dataset contains more than 5000 measurements, including 1024 JLab DIS points. The new data provides precision data on an $x$ range comparable to SLAC measurements, with an expanded $Q^2$ coverage that helps determine which part of the observed low-energy structure is leading twist and which part is power suppressed.

In the CJ framework the quantities called ``higher twist'' are phenomenological fitted $1/Q^2$ corrections. They stabilize the extraction of the leading-twist PDFs and the off-shell deformation of bound nucleons, but they should not be interpreted as a direct measurement of genuine twist-4 multiparton matrix elements. The distinction becomes especially relevant when proton and neutron-sensitive data are combined: different assumptions about the $Q^2$ dependence and isospin structure of the power corrections can otherwise be traded against the neutron-to-proton structure-function ratio and the nuclear correction.

In this contribution I summarize the main results presented at DIS2026 and emphasize several points that became clearer in the subsequent discussion of the CJ26 analysis. In particular, I explain the role of the JLab 6 and 12~GeV data, the motivation for the two HT implementations used to estimate a theoretical systematic uncertainty, and the meaning of the uncertainty bands quoted for the final CJ26 PDFs.

\section{The CJ strategy at large $x$}

The CJ philosophy is to retain as much of the experimental information as possible and to model the physics that becomes important when the standard leading-twist approximation is no longer sufficient. This is complementary to global analyses that impose more restrictive $W^2$ or $Q^2$ cuts. The latter strategy provides a clean perturbative domain, while the CJ strategy uses the low-energy data to obtain additional information on both partonic and nuclear dynamics.

The leading-twist PDFs are parametrized at an initial scale $Q_0^2=1.69~\mathrm{GeV}^2$ and evolved with NLO DGLAP evolution. The deuteron is treated in the weak-binding approximation, with nucleon motion and binding included through the nuclear convolution. The structure of a bound nucleon is further modified by an off-shell deformation,
\begin{equation}
 \phi(x,Q^2,p^2)=\phi^{\rm free}(x,Q^2)
 \left[1+\frac{p^2-M^2}{M^2}\,\delta f(x)\right],
 \label{eq:offshell}
\end{equation}
where $\delta f(x)$ is fitted simultaneously with the PDFs. 
%This construction is essential for using deuteron data to separate $u$ and $d$ distributions.
%
The off-shell function %is not an arbitrary correction added after the PDF fit. It 
is part of the leading-twist nuclear description and is therefore correlated with the neutron structure function. This correlation is one reason why the range of $Q^2$ covered by the data matters so much. At low $Q^2$, an off-shell modification can resemble a power correction, whereas the different $Q^2$ dependence of the two effects become easier to distinguish once data at substantially different scales and similar $x$ values are available.

The other important ingredient is the treatment of target-mass and residual power corrections. Target-mass corrections are kinematic effects of order $M^2/Q^2$ that can are calculated using an approximation of the Georgi-Politzer method \cite{Accardi:2026hdv}. Remaining power corrections are parametrized in either multiplicative or additive form,
\begin{align}
 F_{2,N}^{\rm mult}(x,Q^2)
 &=F_{2,N}^{\rm TMC}(x,Q^2)
   \left[1+\frac{C_N(x)}{Q^2}\right],
 \label{eq:mht}\\
 F_{2,N}^{\rm add}(x,Q^2)
 &=F_{2,N}^{\rm TMC}(x,Q^2)
   +\frac{H_N(x)}{Q^2},
 \label{eq:aht}
\end{align}
with $N=p,n$. 
%The two forms are mathematically different parameterizations of the same broad class of phenomenological power corrections. 
There is no compelling first-principles reason to select one representation over the other. %in the kinematic region of interest.
%
%This observation has a practical consequence that is central to the CJ approach. A multiplicative correction
In fact, a scale-independent, multiplicative correction with $C_p=C_n \equiv C$ is equivalent to an additive correction
\begin{equation}
 \widetilde H_N(x,Q^2)=F_{2,N}^{\rm TMC}(x,Q^2)\,C(x)
\end{equation}
and thus $\widetilde H_N(x,Q^2)$ explicitly inherits both the $Q^2$ evolution and the proton-neutron dependence from  the leading-twist structure function in contrast to the pure additive $H_N(x)$ from \cref{eq:aht}. 

While the precision of the available data is not sufficient to directly constrain the scale behavior of either the $C$ or $H$ coefficients, the CJ22ht study \cite{Cerutti:2025yji} showed that imposing isospin-independent coefficients in either representation introduces a large, spurious implementation dependence into the extracted neutron-to-proton ratio.
CJ26 addresses this inherent implementation ambiguity by fitting proton and neutron HT functions independently and by repeating the global fit with both additive and multiplicative implementations. The spread between the two fits is then used as an estimate of the associated ``epistemic'' uncertainty. 

\section{What is new in CJ26?}

The CJ26 dataset contains 5054 fitted points before normalization contributions, of which 1024 come from Jefferson Lab, substantially more than in the CJ22ht analysis \cite{Cerutti:2025yji}. 
%As shown in Figure~\ref{fig:coverage}, 
The 6~GeV measurements provide a dense set of high-precision points at moderate and large $x$, but they occupy a relatively narrow range in $Q^2$. They are consequently especially sensitive to the size and shape of the power corrections. The 12~GeV data extend the accessible $Q^2$ range while retaining large-$x$ reach. They therefore provide the lever arm needed to distinguish a change in the leading-twist partonic structure from a genuine $1/Q^2$ effect. 

% \begin{figure}[t]
% \centering
% % \includegraphics[width=\textwidth]{CJ26_fig1.pdf}
% \fbox{\parbox[c][0.16\textheight][c]{0.94\textwidth}{\centering
% Suggested Fig.~1: reproduce/adapt CJ26 paper Fig.~1, showing the $x$--$Q^2$ coverage of the DIS, vector-boson, and jet data.}}
% \caption{Suggested representative figure: kinematic coverage of the CJ26 global dataset. The main point to emphasize is the extension of the DIS data into the large-$x$, low-$Q^2$ region by the JLab program, together with the additional $Q^2$ leverage provided by the 12~GeV measurements.}
% \label{fig:coverage}
% \end{figure}

%This distinction is visible already in the quality of the global fit. The overall description is good, with the remaining tensions concentrated in particular datasets and kinematic regions rather than appearing as a global failure of the framework. 
An important part of obtaining this result is the treatment of experimental correlations. Whenever the experimental information is available, correlated systematic uncertainties are included explicitly in the fit rather than simply added in quadrature to the statistical errors.
This is important because if a systematic uncertainty is correlated across many data points, treating it as independent point by point changes both the effective weight of the data and the pattern of residuals. The CJ26 study finds clear examples in the SLAC and JLab data where removing the correlated shifts and adding their uncertainties in quadrature produces noticeably distorted pull distributions \cite{Accardi:2026hdv}. This makes a complete specification of systematic correlations in published experimental results an important ingredient in precision global fitting.

Finally, with this substantially larger DIS data set, CJ26 released some of the PDF parameters that were previously kept at their best-fit value, and enlarged the parametrization of the off-shell and HT functions.

\section{JLab and the separation of leading and higher twist}

The impact of the new data is most clearly seen by comparing the nominal CJ26 fit with a baseline fit performed with the same theoretical and parametrization choices but without JLab data. %This comparison isolates the effect of the new measurements rather than mixing it with changes in the fit setup.

\begin{figure}[t]
\centering
    \includegraphics[width=0.9\linewidth]{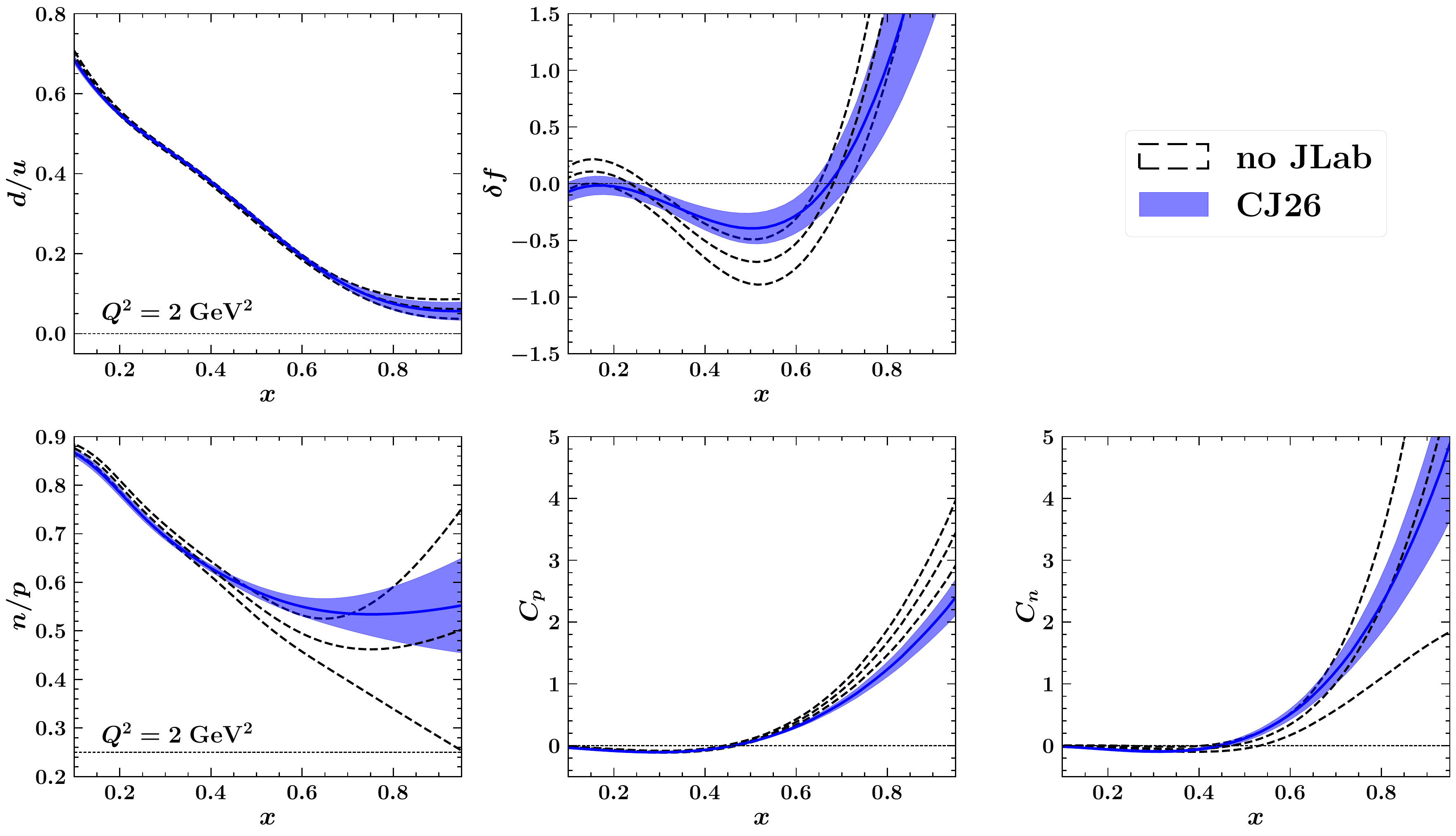}
    \caption{Impact of the full JLab dataset on the extracted large-$x$ quantities. The largest gains are in the power corrections and nuclear quantities. The $d/u$ is comparatively stable, yet with reduced uncertainty \cite{Accardi:2026hdv}. }
\label{fig:jlabimpact}
\end{figure}

The main result is that the JLab data substantially reduce the uncertainty on several quantities that were previously poorly separated. In particular, the uncertainty of the neutron HT correction and of the neutron-to-proton structure-function ratio decreases markedly. The $d/u$ ratio is more stable in central value, but its relative uncertainty is reduced by roughly 10\% up to $x\simeq0.6$ and by several percent out to $x\simeq0.8$. At still larger $x$, the result is increasingly controlled by extrapolation and should not be interpreted as a direct experimental constraint.

The off-shell function is also affected. With the more flexible third-order polynomial used in CJ26, the fit finds a nontrivial shape that becomes negative around $x\simeq0.2$--0.4, approaches zero at intermediate $x$, and turns upward at larger $x$. This differs qualitatively from the concave shape obtained in the earlier CJ22ht analysis. The change is not evidence for a sudden change in the underlying nuclear physics; rather, it illustrates the danger of using an unnecessarily restrictive parametrization when new data begin to constrain a previously weakly determined function. In CJ26, fourth-order fits give compatible results over the experimentally constrained region, supporting the use of the third-order form.

To summarize, the JLab measurements do more than simply reduce the PDF errors. They determine how much of the low-$Q^2$ behavior must be assigned to power corrections and how much can be attributed to leading-twist nuclear dynamics. The 6~GeV measurements are particularly effective at constraining the power corrections, while the 12~GeV data extend the $Q^2$ lever arm and thereby improve the separation from the off-shell contribution. This is the sense in which the CJ26 result provides a more robust, rather than perfectly model-independent, separation of effects.

The HT and off-shell contributions are not completely disentangled by the present data, and the CJ26 analysis does not claim that they are. Rather, the combination of the full 6~GeV sample, the new 12~GeV measurements, and explicit tests of implementation and parametrization dependence substantially reduces the ambiguity compared with earlier CJ fits. Future 12 GeV data from JLab and data at still higher $Q^2$ such as from the EIC are expected to provide an even cleaner separation and to substantially improve the precision of the $d/u$ extraction.

%\section{Why proton and neutron higher twists are fitted separately}
\section{Isospin dependence of power corrections}

A recurring question is whether the difference between the fitted proton and neutron HT functions can be interpreted as isospin violation in genuine twist-4 matrix elements. It cannot: the CJ HT terms are phenomenological $1/Q^2$ corrections that absorb several sources of residual power dependence, including genuine multiparton correlations, approximations in target-mass and nuclear corrections, and threshold effects. In particular, fitted HT terms can absorb part of the missing higher-order perturbative contribution \cite{Harland-Lang:2025wvm,Ball:2025xtj} and thereby stabilize the NLO leading-twist extraction.

Even with isospin-symmetric QCD dynamics, proton and neutron HT terms can differ phenomenologically because their leading-twist structure functions have different $x$ dependence. A multiplicative correction therefore corresponds to different additive corrections for the two nucleons; approximations in nuclear mass corrections can likewise affect them differently. CJ26 consequently fits proton and neutron HT independently, reducing the risk that an imposed isospin relation in the HT sector is compensated by changes in the neutron PDF or off-shell function.

The fitted $H_p-H_n$ difference provides information about the phenomenological power corrections required by the global fit, not a direct measurement of isospin violation in twist-4 operators. Yet, this differences helps explain why $F_{2,n}/F_{2,p}$ and $d/u$ can behave differently at large $x$: the former is directly sensitive to the proton and neutron power corrections, while the latter is a leading-twist PDF ratio. In CJ26 the enhanced large-$x$ tail of $F_{2,n}/F_{2,p}$ is substantially driven by the different fitted proton and neutron power corrections, whereas $d/u$ remains comparatively stable.

\section{Estimating the CJ26 uncertainty}

The uncertainty bands shown in CJ26 contain two conceptually different ingredients. First, each fit is analyzed with a Hessian uncertainty based on a tolerance $T^2=2.71$. This tolerance is larger than one because the global dataset contains tensions and because the likelihood can deviate from a simple Gaussian form, especially along weakly constrained directions. The tolerance should therefore not be interpreted literally as a universal two-standard-deviation probability interval.

Second, CJ26 performs two fits using the additive and multiplicative HT implementations. The minima of the two fits differ by about $\Delta\chi^2=31$. Increasing the Hessian tolerance within either fit does not automatically cover this difference because the two implementations correspond to distinct solutions in the large-$x$ \textit{functional} space. The nominal CJ26 uncertainty is therefore taken as the envelope of the Hessian bands from the two implementations, combining parameter uncertainty and the implementation uncertainty of the phenomenological power corrections.

\begin{figure}[t]
\centering
    \includegraphics[width=0.9\linewidth]{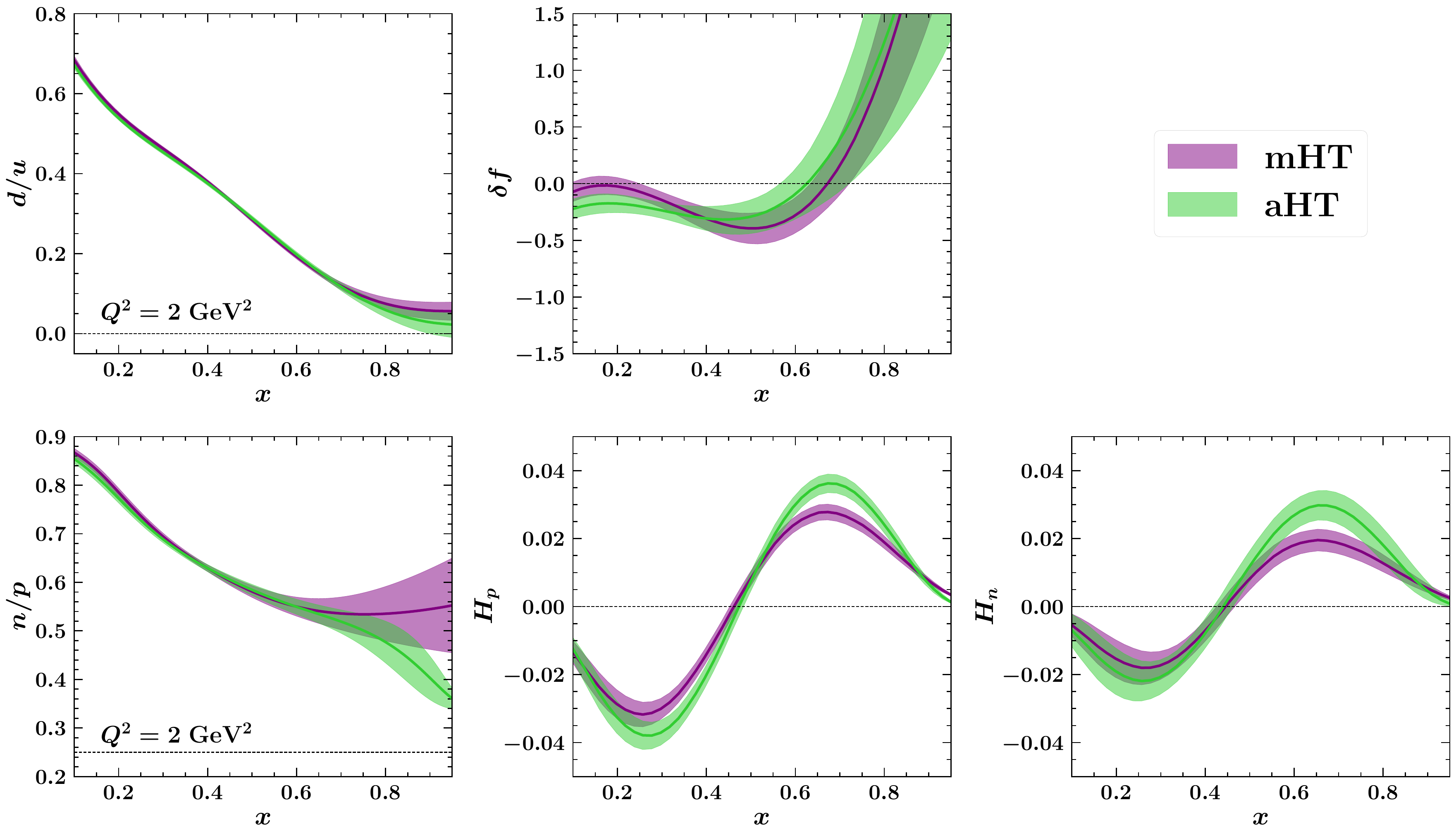}
    \caption{Stability of the extracted quantities against the implementation of higher-twist corrections. The envelope of the two fits is the CJ26 prescription for incorporating this source of epistemic uncertainty.}
\label{fig:htsystematics}
\end{figure}

The important observation is that the leading-twist quantities are remarkably stable under this variation. The $d/u$ ratio and the off-shell function agree between the two implementations over the experimentally constrained region. The neutron-to-proton structure-function ratio is likewise stable up to approximately $x=0.8$--0.85, with differences becoming more visible only where the data no longer constrain the extrapolation. This stability is one of the principal checks supporting the CJ26 extraction.

The remaining implementation dependence identifies the region in which the present data do not fully determine the functional form of the power corrections. In that sense, comparing the two HT implementations is not merely a technical exercise; it provides a concrete diagnostic of where the global fit is data driven and where model assumptions become increasingly important.

\section{Results and perspective}
CJ26 shows that the full JLab data set adds more than statistical precision. The 6 GeV measurements constrain the power corrections, while the 12 GeV data extend the large-$x$ $Q^2$ lever arm needed to separate power-suppressed effects from leading-twist nuclear dynamics. The extracted $d/u$, $F_{2,n}/F_{2,p}$, and off-shell function are consequently more tightly constrained and remain stable under the two HT implementations over the experimentally relevant region.

Several measurements can further test this framework. BONuS12 tagged-neutron data will directly constraint the neutron structure function. New 12 GeV deuteron measurements will extend the available $Q^2$ coverage. Parity-violating DIS at SoLID will at the same time reduce reliance on nuclear targets and validate the nuclear correction framework. Higher-$Q^2$ measurements at the EIC will provide the additional leverage needed for cleanly disentangling LT and HT dynamics.

The broader lesson is methodological: at large $x$, accurate PDF extraction requires the simultaneous treatment of target-mass, power-suppressed, and nuclear effects, together with a 
%correlated experimental uncertainties and a 
careful study of their implementation systematics. The stability of the CJ26 PDFs under HT implementation variation provides a useful foundation for the next generation of large-$x$ global analyses not only at NLO as in the work discussed here  \cite{Accardi:2026hdv} but also at any higher order in perturbation theory \cite{Harland-Lang:2025wvm}.

\section*{Acknowledgments}

This work was supported in part by the U.S. Department of Energy under contracts DE-SC0008724 and DE-AC05-06OR23177. The initial draft was generated with the aid of ChatGPT-6.1 Luna using the presenter's slides, the CJ26 paper, and internal collaboration memos as source material. The authors reworked, validated, and take full responsibility for the final output.

\bibliographystyle{JHEP}
\bibliography{CJ}

\end{document}